\mag=\magstephalf
%\mag=\magstep1
\pageno=1
\input amstex
%\baselineskip = 0.8 true cm
\documentstyle{amsppt}
\TagsOnRight
\NoRunningHeads

\pagewidth{16.5 truecm}
\pageheight{23.0 truecm}
\vcorrection{-1.0cm}
\hcorrection{-0.5cm}
%\advance\vsize by -\voffset
%\advance\hsize by -\voffset
%\baselineskip = 1.0 true cm
\nologo

\NoBlackBoxes
\font\twobf=cmbx12

\define \ee{\roman e}
\define \dd{\roman d}
\define \tr{\roman {tr}}

\define \CC{{\Bbb C}}

\define \diag{\roman {diag}}
\define \off{\roman {off}}
\define \tu{{\tilde u}}
\define \tv{{\tilde v}}

%\define \tvskip{\vskip 0.5 cm}
\define \tvskip{\vskip 1.0 cm}
\define\ce#1{\lceil#1\rceil}
\define\dg#1{(d^{\circ}\geq#1)}
\define\Dg#1#2{(d^{\circ}(#1)\geq#2)}
\define\dint{\dsize\int}
\def\fp{\flushpar}

\define\s#1{\sigma_{#1}}
\define\tp#1{\negthinspace\left.\ ^t#1\right.}
\define\mrm#1{\text{\rm#1}}
\define\lr#1{^{\sssize\left(#1\right)}}

\redefine\qed{\hbox{\vrule height6pt width3pt depth0pt}}
\font\Large=cmr10 scaled \magstep5

{\centerline{\bf{Soliton Solutions of Korteweg-de
Vries Equations}}}

{\centerline{\bf{and}}}

{\centerline{\bf{
Hyperelliptic Sigma Functions}}}

\author
Shigeki MATSUTANI${}^0$%\footnote
\endauthor
\affil
8-21-1 Higashi-Linkan Sagamihara 228-0811 Japan
\endaffil
 \endtopmatter

\footnotetext{e-mail:RXB01142\@nifty.ne.jp}

\document
%\baselineskip= 0.8 true cm

\centerline{\twobf Abstract }\tvskip

Soliton Solutions of  Korteweg-de Vries (KdV) were constructed
for  given degenerate curves $y^2 = (x-c)P(x)^2$ in terms
of hyperelliptic sigma functions and explicit Abelian integrals.
Connection between sigma functions and tau function were also presented.

%\subheading{PACS numbers}:

%\newpage
\vskip 1 cm
{\centerline{\bf{\S 1. Introduction}}}
\vskip 0.5 cm

The modern soliton theories [DJKM, SN, SS], which were developed in ending
of last century, are known as the infinite dimensional analysis and gave us
fruitful and beautiful results, {\it e.g.}, relations of soliton equations
to universal Grassmannian manifold, Pl\"ucker embedding,
infinite dimensional Lie algebra, loop algebra, representation theories,
Schur functions, Young diagram, and so on.
They stemmed from an investigation of the B\"acklund transformations
among the soliton solutions,  which are expressed by
hyperbolic functions [DJKM, SN, SS]. By primitive considerations, the
soliton solutions are related to certain degenerate
algebraic curves and thus the B\"acklund transformation can be regarded as
transformation among certain  degenerate
curves of different genera  [DJKM, SN, SS].

As the modern soliton theories  are based on the abstract theory,
theories on the Abelian functions established in nineteenth century are
very concrete [B1-3, Kl].
They are of given concrete algebraic curves and
of Abelian functions and differential equations there.
By fixing a hyperelliptic curve,
Klein defined the hyperelliptic sigma function,
a hyperelliptic version of Weierstrass
sigma function for elliptic curve [Kl].
The sigma function is a well-tuned theta function and
brings us fruitful information of the curve.
In terms of the hyperelliptic sigma functions and bilinear
differential operator, Baker discovered the Korteweg-de Vries hierarchy,
Kadomtsev-Petviashvili equation  and gave
their periodic solutions without ambiguous parameters [B1-3, Ma1].

As the new century began, I believe that we should
connect both established theories from a novel point of view.
Recently the theories in nineteenth century has
been re-evaluated in various fields from similar viewpoint
[BEL1-5,CEEK, EE, EEL, EEP, Ma1,2, N].
For example,
Buchstaber, Enolskii and Leykin connected the hyperelliptic
sigma functions with the Paffian in [BEL3] and
the Schur-Weierstrass polynomials in [BEL4].
The purpose of this article is also to give a step to a
unification of both theories.

In this article, we will focus on the soliton solutions of
 Korteweg-de Vries equation in terms of
the hyperelliptic sigma functions over degenerate hyperelliptic
curves. It is needless to say that the soliton solutions
played important roles in developments of soliton theory
and the investigations of the mathematical structure
of the soliton solutions are well-established  [H1,2, GGKM, L, DKJM].
In fact the similar studies to this article
have been already done by Its and Matveev  [BBEIM, I, IM1].
They showed an explicit
connection between so-called Hirota's tau functions of the soliton solutions
and Riemannian theta functions in terms of the normalized differential
of the first kind.
Their works [BBEIM, I, IM1, 2] also influenced the
development of the modern soliton theory.

As in [BEL1,2, Kr1, Kr2, IM1,2],
 solutions space of KdV equation is very fruitful
and the soliton solutions are in very special locus of the space,
I believe that in order to know the solutions space of the KdV equation,
it is very important to recognize again
how special the soliton solution is.

Further recent studies [BEL1-5,CEEK, EE, EEL, EEP, Ma1,2, N]
shows that the unnormalized differentials (one-forms) are more important
than the normalized ones used in [BBEIM, I, IM1].
Thus we believe that we should re-consider these
studies of forerunners in terms of more clear words.

In other words, this article should be regarded as
a revise of the works of Its and Matveev [I, IM1].
For example
they found the nice differentials of the
first kind (3-6), which is a key of this study,
and obtained the most of results as this article.
However from viewpoint of algorithmic investigations,
their approach is a little bit heuristic.
Further it is neither clear why the soliton solutions
in different genera are connected whereas those of
more general hyperelliptic curves are not found
 [Kr1, 2, BBEIM].
They neither commented on the importance of the
existence of (3-6) nor the specialty of the soliton
solutions.

In this article, by employing older fashion as in nineteenth century,
we start with a concrete degenerate curve $ y^2 = P(x)^2 x$ and
reconstruct the soliton solutions after performing some explicit
Abelian integrations.
By dealing with the unnormalized differentials of the first kind,
we find  nice differentials (one-form) $d v$'s in (3-6)
over the special curve $y^2=P(x)^2 x$, which consists only of the data
of zero of $P(x)$ and does not depend upon other global structure of
the curves. The existence of the differentials characterizes
the soliton solutions in the solution space of the KdV equation
and a key of the fact that the soliton solutions are very simple.
As shown in lemma 3-6,
the existence of the differentials simplifies the Abelian integrations.
For examples, the components of the integration matrices do not depend upon
the genus of curve.  Accordingly we can easily find  connections among
such special curves with different genera.
However from the study of hyperelliptic functions in nineteenth century,
such  differentials do not exist for general hyperelliptic curves as
shown in Remark 3-10.  Our concrete computations and the theories in nineteenth
century enable us to recognize that such connections
among the curves with different genera might have practical meanings
 {\it only} for such special cases.
In other words, as I point out there, the B\"acklund transformation
might be effective {\it only} for the degenerate curves associated with
the soliton solutions.

\vskip 0.5 cm

In order to unify the both theories in nineteenth and twentieth centuries,
we should also recognize the difference between them.
Thus I considered this problem.
After submitting this article,
I knew the works of Its and Matveev [BBEIM, I, IM1]
and of Edelstein, G\'omez-Reino and Mari\~no EGM] and revised this.
Edelstein, G\'omez-Reino and Mari\~no also dealt with soliton solutions
based upon the formula in [BBEIM] in the context of the gauge field theory
after considering the hyperelliptic sigma functions.
However as their purpose is not to unify the both theories,
their treatments are not enough from our viewpoint.
Even after knowing the work [I, IM1, EGM],
 I believe that  concrete computations of the soliton solutions in the
framework of sigma function theory has very important meanings
in this stage.
For example, as the coordinate systems (conventions) in the Jacobian
are different in both theories, {\it i.e.}, $\bold u$ to that of the
nineteenth and $\bold t$ to that of the twentieth centuries,
I connect them in (3-19) in terms of a matrix (3-3) for the case
of the soliton solutions. Further as the coefficient of these coordinate
systems in the theta functions looks different between
the theories in 1970's  [Kr1, 2, IM1, 2]
 and the definition of Klein, it turns out that
they are connected by Legendre relation (2-23) and (3-33).
As mention it in Remark 3-10,
comparison between one-form (3-6) and (3-3) appearing in Baker's
investigation [Ba2] shows how special the soliton solutions are.

\vskip 0.5 cm

{\centerline{\bf{\S 2. Preliminary of Baker's
Hyperelliptic Sigma Functions}}}

\vskip 0.5 cm

In this section, we will review the hyperelliptic sigma function.
In this article, we will mainly use the
conventions of \^Onishi [\^O1]. As
there is a good self-contained paper on
theories of hyperelliptic sigma functions [BEL2]
besides [B1,2,3],
we will give their notations, definitions and
propositions without explanations and proves.

We denote the set of complex number by $\Bbb C$ and
the set of integers by $\Bbb Z$.

\proclaim{\fp Notation 2-1}\it
We deal with a hyperelliptic curve $\tilde C_g$  of genus $g$
$(g>0)$ given by an affine equation,
$$ \split
   y^2 &= f(x) \\  &= \lambda_{2g+1} x^{2g+1} +
\lambda_{2g} x^{2g}+\cdots  +\lambda_2 x^2
+\lambda_1 x+\lambda_0  \\
     &=P(x)Q(x)\\
\endsplit  \tag 2-1 $$
where $\lambda_{2g+1}\equiv1$ and $\lambda_j$'s and
$$
\split
        Q(x) &= (x-c_1)(x-c_2)\cdots(x-c_g)(x-c),\\
	P(x) &= (x-a_1)(x-a_2)\cdots(x-a_g) \\
             &= x^g + \mu_{g-1} x^{g-1}+\cdots
            +\mu_2 x^2+\mu_1 x + \mu_0,
\endsplit \tag 2-2
$$ $c_j$'s, $a_j$'s, $c$  and $\mu_j$'s
are  complex values.
\endproclaim

We will express a point $\mrm P$ in the curve
by the affine coordinate $(x,y)$ if it is not the
infinity point.
\vskip 0.5 cm

\proclaim{\fp Definition 2-2 [B1 p.195, B2 p.314, B3
p.137, BEL2 Chapter 2, \^O1 p.385-6]}\it

 \roster

\item  Let us denote the homology of a hyperelliptic
curve $\tilde C_g $ by
$$
\roman{H}_1(\tilde C_g, \Bbb Z)
  =\bigoplus_{j=1}^g\Bbb Z\alpha_{j}
   \oplus\bigoplus_{j=1}^g\Bbb Z\beta_{j},
 \tag 2-3
$$
where these intersections are given as
$[\alpha_i, \alpha_j]=0$, $[\beta_i, \beta_j]=0$ and
$[\alpha_i, \beta_j]=\delta_{i,j}$.
These $\alpha$'s and $\beta$'s are given as
illustrated in Fig.1 for the case of genus five.
There we construct the hyperelliptic Riemannian surface using twin
Riemannian spheres with cuts.

\item The unnormalized differentials of the first kind are
defined by,
$$   d u_1 := \frac{ d x}{2y}, \quad
      d u_2 :=  \frac{x d x}{2y}, \quad \cdots, \quad
     d u_g :=\frac{x^{g-1} d x}{2 y}.
      \tag 2-4
$$

\item The unnormalized differentials of the second kind are
 defined by,
$$   d \tu_1 := \frac{x^g d x}{2y}, \quad
      d \tu_2 :=  \frac{x^{g+1} d x}{2y}, \quad \cdots,
 \quad
     d \tu_g :=\frac{x^{2g-1} d x}{2 y},
      \tag 2-5
$$
and $d \bold r := (d r_1, d r_2, \cdots, d r_g)$,
$$
     (d \bold r):=\Lambda \pmatrix d \bold u \\ d \tilde{\bold u}
 \endpmatrix,
     \tag 2-6
$$
where $\Lambda$ is $2g \times g$ matrix defined by
%{\eightpoint
$$
\split
	\Lambda& =
\left(\matrix 0 & \lambda_3 & 2 \lambda_4 &
            3 \lambda_5 & \cdots &
          (g-1)\lambda_{g+1}& g \lambda_{g+2
          }& (g+1)\lambda_{g+3}\\
        \  & 0 &  \lambda_5 & 2\lambda_6 & \cdots &
            (g-2)\lambda_{g+2}& (g-1) \lambda_{g+3}&  g
            \lambda_{g+4}\\
    \ & \ &  0         &  \lambda_7 & \cdots &
            (g-3)\lambda_{g+3}& (g-2) \lambda_{g+4}
               & (g-1)\lambda_{g+5}\\
   \ & \ &  \        & \ & \ddots &
            \vdots & \vdots& \vdots\\
 \ &  & \text{\Large 0}        &  \ & \ &
            \lambda_{2g-2} &2 \lambda_{2g-1}& 3\lambda_{2g+1}\\
 \ &  & \        &  \ & \ &
          0  & \lambda_{2g+1}& 0  \endmatrix\right.\\
&\qquad \qquad
\left.\matrix \cdots & (2g-3)\lambda_{2g-1} & (2g-2)\lambda_{2g}
             & (2g-1) \lambda_{2g+1}\\
          \cdots & (2g-4)\lambda_{2g} & (2g-3)\lambda_{2g+1}
             & 0                    \\
          \cdots & (2g-5)\lambda_{2g+1} &                 0
             &                     \\
\cdots & 0 &
             &                  \ \\
     \  & \ &
             &                     \\
\ & \ &  \text{\Large 0}
             &                  \ \\
\ & \ &  \
             &                  \ \endmatrix\right).
\endsplit \tag 2-7
$$
%}

\item The unnormalized period matrices are defined by,
$$    \pmb{\omega}':=\left[\int_{\alpha_{j}}d u_{i}\right],
\quad
      \pmb{\omega}'':=\left[\int_{\beta_{j}}d u_{i}\right],
 \quad
    \pmb{\omega}:=\left[\matrix \pmb{\omega}' \\ \pmb{\omega}''
     \endmatrix\right].
  \tag 2-8
$$

\item The normalized period matrices are given by,
$$    \ ^t\left[\matrix d \hat u_{1} & \cdots
   d \hat u_{g}
        \endmatrix\right]
       :={\pmb{\omega}'}^{-1}  \ ^t\left[\matrix
          d u_{1} & \cdots
   d u_{g}\endmatrix\right] ,\quad
   \pmb \tau:={\pmb{\omega}'}^{-1}\pmb{\omega}'',
   \quad
    \hat{\pmb{\omega}}:=\left[\matrix 1_g \\ \pmb \tau
     \endmatrix\right].
       \tag 2-9
$$

\item The complete hyperelliptic integrals of
the second kind are given  as
$$      \eta':=\left[\dint_{\alpha_{j}}d r_{i}\right],
\quad
         \eta'':=\left[\dint_{\beta_{j}}d r_{i}\right] .
       \tag 2-10
$$

\item
By defining the Abelian map for $g$-th symmetric product
of the curve $\tilde C_g$ and  for points $\{ Q_i\}_{i=1,\cdots,g}$
in the curve,
$$       \hat w: \roman{Sym}^g(\tilde C_g) \longrightarrow
        \Bbb C^g, \quad
      \left( \hat w_k(Q_i):=\sum_{i=1}^g \int_\infty^{Q_i}
      d  \hat u_k \right),
$$ $$ w:\roman{Sym}^g(\tilde C_g) \longrightarrow \Bbb C^g, \quad
      \left( w_k(Q_i):= \sum_{i=1}^g
       \int_\infty^{Q_i} d u_k \right),
      \tag 2-11
$$
the Jacobi varieties $\hat{\Cal J_g}$ and $\Cal J_g$
are defined as complex torus,
$$   \hat{\Cal J_g} := \Bbb C^g /\hat{ \pmb{\Lambda}} ,
   \quad {\Cal J_g} := \Bbb C^g /{ \pmb{\Lambda}} .
     \tag 2-12
$$
Here $\hat{ \pmb{\Lambda}}$ $({ \pmb{\Lambda}})$  is a
lattice generated by
$\hat{\pmb{\omega}}$ $({\pmb{\omega}})$.

\item We defined the theta function over $\Bbb C^g$
characterized by $\hat{ \pmb{\Lambda}}$ or ${\pmb{\tau}}$,
$$\vartheta\negthinspace\left[\matrix a \\ b \endmatrix\right]
     (z; \pmb \tau)
    :=\sum_{n \in \Bbb Z^g} \exp \left[2\pi \sqrt{-1}\left\{
    \dfrac 12 \ ^t\negthinspace (n+a)\pmb \tau(n+a)
    + \ ^t\negthinspace (n+a)(z+b)\right\}\right],
     \tag 2-13
$$
for $g$-dimensional vectors $a$ and $b$.

\endroster
\endproclaim

\tvskip

\vskip 0.5 cm
\proclaim {Definition 2-3 ($\wp$-function, Baker)
[B1, B2 p.336, p.358, p.370, BEL2 p.35, \^O1 p.386-7] }

\it
\roster
The coordinate in $\Bbb C^g$ for
 points $(x_i,y_i)_{i=1,\cdots,g}$
of the curve $y^2 = f(x)$ is given by,
$$
  u_j :=\sum_{i=1}^g\int^{(x_i,y_i)}_\infty d u_j .
    \tag 2-14
$$

\item Using the coordinate $u_j$, sigma functions,
which is a holomorphic
function over $\Bbb C^g$, is defined by
$$ \sigma(u)=\sigma(u;\tilde C_g):
  =\ \gamma\roman{exp}(-\dfrac{1}{2}\ ^t\ u
  \pmb{\eta}'{\pmb{\omega}'}^{-1}u)
  \theta(u) .
     \tag 2-15
$$
where
$$
\theta(u):=\vartheta\negthinspace
  \left[\matrix \delta'' \\ \delta' \endmatrix\right]
  ({\pmb{\omega}'}^{-1}u ;\pmb \tau),\tag 2-16
$$
and
$$
 \delta' =\ ^t\left[\matrix \dfrac {g}{2} & \dfrac{g-1}{2}
       & \cdots
      & \dfrac {1}{2}\endmatrix\right],
   \quad \delta''=\ ^t\left[\matrix \dfrac{1}{2} & \cdots
& \dfrac{1}{2}
   \endmatrix\right],
     \tag 2-17
$$
$\gamma$ is a certain constant factor.

\item
In terms of these functions,
$\wp$-functions are defined by
$$   \wp_{\mu\nu}(u)=-\dfrac{\partial^2}{\partial
   u_\mu\partial u_\nu}
   \log \sigma(u) ,
         \tag 2-18
$$
$$
  \tilde\wp_{\mu\nu}(u)=-\dfrac{\partial^2}{\partial
   u_\mu\partial u_\nu}
   \log \theta(u).
         \tag 2-19
$$

\endroster

\endproclaim

It is easy to find the difference between $\wp$
 and $\tilde \wp$ as follows.

\proclaim{\fp Lemma 2-4 }\it

$$
	\tilde\wp_{ij}(u) = \wp_{ij}(u)
 - C_{ij}, \tag 2-20
$$
where
$$
	C_{ij}=( \pmb{\eta}'{\pmb{\omega}'}^{-1})_{ij}.
     \tag  2-21
$$
\endproclaim

\proclaim{\fp Proposition 2-5 [B3, BEL2 p.52,53, Ma1]}\it

$\wp_{\mu\nu\rho}:=\partial \wp_{\mu\nu}(u)
  /\partial u_\rho$ and
$\wp_{\mu\nu\rho\lambda}:=\partial^2
 \wp_{\mu\nu}(u) /\partial u_\mu \partial u_\nu$.
Then hyperelliptic $U:=(2\wp_{gg}+\lambda_{2g}/6)$
 obeys the Korteweg-de Vries
equations,
$$
	4\frac{\partial U}{\partial u_{g-1}}
        +6 U \frac{\partial U}{\partial u_g}
               + \frac{\partial^3 U }{\partial u_g}
         =0.\tag 2-22
$$

\endproclaim
We should note that this proposition is easily proved
in the framework of sigma function theories [BEL2, Ma1, Ma3].
In fact, Baker already discovered this fact [B3] around 1898
as mentioned in [Ma1];
his way is very simple as explained in [Ma1, Ma3].
However as $\tilde \wp$ and $\wp$ differ just by
the constant matrix from (2-20),
we should mention the works of pioneers [BBEIM, Kr1, IM1, 2];
they directly showed it in terms of  $\tilde \wp$.

\proclaim{\fp Proposition 2-6 [BEL2 p.11 ]}\it

The Legendre relation is given by
$$
  {}^t\omega'\eta''- {}^t\omega''\eta' = 2\pi\sqrt{-1}I_g, \tag 2-23
$$
where $I_g$ is the $g\times g$-unit matrix.

\endproclaim

\vskip 0.5 cm

{\centerline{\bf{\S 3. Soliton Solutions of
 Korteweg-de Vries Equations}}}

\vskip 0.5 cm

In this section, we will consider a degenerate
curve which is connected with the so-called
soliton solution. For a degenerate curve, the differentials
(2-4) and (2-5) become singular as we show later,  but the
sigma function can be defined as a limit of non-singular
curves even though it needs some regularizations.
 In fact, we will show that the sigma function
over a certain curve can be associated with tau function
after regularizations.

\proclaim{\fp Degenerate Curves 3-1}\it
We deal with a degenerate hyperelliptic curve $C_g$  of genus $g$
$(g>0)$ given by an affine equation,
$$
   y^2 =P(x)^2 x , \tag 3-1
$$
i.e., $Q(x) = P(x)x$ in (2-2), or
$c \equiv0,\	c_1\equiv a_1, \	c_2\equiv a_2,
	\ \cdots, \ 	c_g\equiv a_g.
$ in (2-1) and (2-2).
\endproclaim

For later convenience, we introduce parameters
$(k_i)_{i=1,\cdots,g}$, $k_i = \sqrt{a_i}$.
Here we note that since the affine equation has symmetries,
{\it i.e.}, translation, dilatation, inversion,
in a projective space, it should be regarded as
a representation element in an equivalent relations.
In this article, I use the representation for simplicity.

\proclaim{\fp Definition 3-2}\it
\roster
\item For $g$-th polynomial $P(x)$ in (2-2), we define,
$$
\split
\pi_i(x) &:= \frac{P(x)}{x-a_i}\\
        &=\chi_{i,g-1}x^{g-1} +\chi_{i,g-2} x^{g-2}
            +\cdots+\chi_{i,1}+\chi_{i,0}.\\
\endsplit \tag 3-2
$$

\item We will introduce $g\times g$-matrices
$$
 W := \pmatrix
     \chi_{1,0} & \chi_{1,1} & \cdots & \chi_{1,g-1}  \\
      \chi_{2,0} & \chi_{2,1} & \cdots & \chi_{2,g-1}  \\
   \vdots & \vdots & \ddots & \vdots  \\
    \chi_{g,0} & \chi_{g,1} & \cdots & \chi_{g,g-1}
     \endpmatrix,\quad
M:= \pmatrix
      1         & \ & \ & \ &\  \\
      \mu_{g-1} & 1 & \ &\text{\Large 0}  & \  \\
   \mu_{g-2} &\mu_{g-1}     & 1      &  & \ \\
    \vdots& \vdots          & \vdots & \ddots & \\
   \mu_{1} &\mu_{2}     & \cdots      & \mu_{g-1} & 1 \\
     \endpmatrix,
$$
$$
	\Cal P = \pmatrix
     P'(a_1) & \ & \ & \  \\
      \ & P'(a_2)& \ & \   \\
      \ & \ & \ddots   & \   \\
      \ & \ & \ & P'(a_{g})  \endpmatrix,\quad
K(l) := \pmatrix
     k_1^{2g+l-2} &   k_1^{2g+l-4} & \cdots & k_1^l  \\
  k_2^{2g+l-2} &   k_2^{2g+l-4} & \cdots & k_2^l  \\
   \vdots & \vdots & \ddots & \vdots  \\
  k_g^{2g+l-2} &   k_g^{2g+l-4} & \cdots & k_g^l
     \endpmatrix,\tag 3-3
$$
where $P'(x):=d P(x)/d x$ and $\mu$'s are defined in (2-2).
\endroster
\endproclaim

\proclaim{\fp Lemma 3-3}\it

\roster

\item $W_{ig}=^t(1,1,\cdots,1)$ and
$$
          W= K(0) M. \tag 3-4
$$

\item
The unnormalized differentials (2-4) and (2-5) become
given by
$$   d u_i = \frac{ s^{2i-2}ds}{P(s^2)}, \quad
d \tu_i = \frac{ s^{2g+2i-2}ds}{P(s^2)},
      \tag 3-5
$$
where $s := \sqrt{x}$.

\item The other unnormalized differentials
defined by
$$
	d v_i := \sum_{j=1}^g W_{ij} d u_j, \quad
     d \tv_i := \sum_{j=1}^g W_{ij} d \tu_j, \tag 3-6
$$
are expressed by
$$
\split
	(dv_1,dv_2,\cdots, dv_g)
         &= \left( \frac{ds}{(s^2-a_1)}, \frac{ds}{(s^2-a_2)},
           \cdots,  \frac{ds}{(s^2-a_g)}\right) ,\\
	(d\tv_1,d\tv_2,\cdots, d\tv_g)
         &= \left( \frac{s^{2g}ds}{(s^2-a_1)},
          \frac{s^{2g}ds}{(s^2-a_2)},
           \cdots,  \frac{s^{2g}ds}{(s^2-a_g)}\right) .
\endsplit \tag 3-7
$$

\item The inverse matrix of $W$ is given by $W^{-1}={\Cal P}^{-1} V$,
where $V$ is Vandermonde matrix,
$$
	V= \pmatrix 1 & 1 & \cdots & 1 \\
                   a_1 & a_2 & \cdots & a_g \\
                   a_1^2 & a_2^2 & \cdots & a_g^2 \\
                    \vdots& \vdots &       & \vdots \\
                   a_1^{g-1} & a_2^{g-1} & \cdots & a_g^{g-1}
                 \endpmatrix. \tag 3-8
$$

\item The unnormalized differentials of the second kind $dr$'s
are
$$
     d r_{j}= \Lambda \pmatrix W^{-1} & 0 \\ 0 & W^{-1} \endpmatrix
              \pmatrix d \bold v\\ d \tilde{\bold v}\endpmatrix
    . \tag 3-9
$$

\endroster
\endproclaim

\demo{Proof}
(1) Noting the relation $\pi_i(x) (x-a_i) =P(x)$, we have
$$
	\chi_{i,k}=\mu_{k+1} - a_i \chi_{i,k-1}, \tag 3-10
$$
and the initial condition $\chi_{i,g-1}\equiv 1$. Then we have (1).
Noting the properties of the inverse matrix of  Vandermonde matrix,
others are obtained by direct computations.
\qed \enddemo

\proclaim{\fp Remark  3-4}\rm{
\roster

\item For the degenerate curve in (3-1),
$du$'s are no more holomorphic over $C_g$ due to the behavior of $dv_i$.

\item $\dfrac{1}{k_i} dv_i$ is a differential of third
kind {\it i.e.} a differential with two singular points
whose residues are $+1$ and $-1$ respectively.

\item $d v$'s are composed only of the data of a single zero
of $P(x)$ and do not depend upon the $y$ explicitly.
In other words, $d v$'s do not explicitly carry the data
in which curve it is defined.

\roster
}
\endproclaim

\proclaim{\fp Lemma  3-5}\it
\roster

For the contour $\alpha$'s and $\beta$'s as illustrated
in Fig.2 for genus five case,
we have the following results.

\item
$$
\int_{\alpha_j}d v_i = \pi \sqrt{-1} \frac{1}{k_i} \delta_{j,i},
\quad
\int_{\alpha_j}d \tilde v_i = \pi \sqrt{-1} k_i^{2g-1} \delta_{j,i}.
\tag 3-11
$$

\item
$$
\split
\int_{\beta_j}d v_i &=  \frac{1}{k_i}
            \log\left| \frac{(k_i-k_j)}
                  {(k_i+k_j)}\right|,
\\
\int_{\beta_j}d \tilde v_i &=
  \sum_{r=0}^{g-1} \frac{ k_i^{2r} k_j^{2g-2r-1}}{g-r-1/2}
        + k_i^{2g-1}
            \log\left| \frac{(k_i-k_j)}
                  {(k_i+k_j)}\right|.
\endsplit
 \tag 3-12
$$

\endroster
\endproclaim

\demo{Proof}
The contour $\alpha$ contains only one singularity,
$$
\int_{\alpha_j}d v_i=\oint_{s=k_i} \frac{ds}{(s-k_i)(s+k_i)}
                    = 2\pi\sqrt{-1}\frac{1}{2k_i}. \tag 3-13
$$
Similarly we have integration of $d \tilde v$.
On (2), the contour $\beta$ can be restricted
over real line if $a_i$ are real. Real valued integration $(a_j,\infty)$
 gives (3-12) and it is also true even if $a_i$ is not real valued.
The factor 2 in (3-12) comes from the returned path. \qed
\enddemo

Using lemma 3-5, we have $\pmb\omega'$, $\pmb\omega''$,
$\pmb\tau$, and  $\pmb \eta'$ by direct computations following
these definitions, (2,8-10) and (3-6), as next lemma.

\proclaim{\fp Lemma  3-6}\it

\roster

\item
$$
 \pmb\omega'= W^{-1}\left(\frac{\pi\sqrt{-1}}{k_i}
 \delta_{ij} \right),\quad
{\pmb\omega''}= W^{-1}
        \left( \frac{1}{k_i}  \log\left| \frac{(k_i-k_j)}
                  {(k_i+k_j)}\right| \right). \tag 3-14
$$

\item All components of diagonal part of $\pmb\tau$,
which is denoted by
$\pmb\tau_{\diag}$, are diverge $\tau_{ii}=\sqrt{-1}\infty$
and off diagonal part
$\pmb\tau_{\off} :=\pmb\tau-\pmb\tau_{\diag}$
is given by
$$
 \tau_{ij} =\frac{\sqrt{-1}}{\pi}
 \log\left| \frac{(k_i+k_j)}
                  {(k_i-k_j)}\right|,
\quad i\neq j. \tag 3-15
$$

\item The components of $\pmb \eta'$, $\pmb\eta'=(\eta_{i}(k_j))$,
is given by
$$
	\eta_{i}(k) =\frac{ \pi \sqrt{-1}k^{2i-3}}{P'(k^2)}
               \left[\frac{d}{d x}\left(\frac{f(x)}{x^{2i}}
               \right)_+ \right]_{x=k^2}, \tag 3-16
$$
where $()_+$ is the polynomial part of rational function
of $x$.

\item The matrix $C:=(C_{ij})$ in (2-21) is equal to,
$$
	C =\left(\frac{k_j}{ \pi\sqrt{-1}}\eta_i(k_j)\right) W,
        \tag 3-17
$$
and especially $C_{gg}=1$.
\endroster
\endproclaim

Here we note that the origin of the truncated polynomial
in (3-16) is $\Lambda$ matrix in (2-6) {\it i.e.},
$$
         dr_i = \frac{x^i}{2y}d x \frac{d }{dx}
          \left[ \left( \frac{f(x)}{x^{2i}} \right)_+
          \right]. \tag 3-18
$$
As it might be a digression, one might wonder why such truncated polynomial
appears in definition of sigma function.
One of its reasons is to concentrate the singularities of the differential
of the second kind at the infinity point; a certain residual integral at
the infinity point removes the excess part.
As in soliton theory, we encounter a truncated differential
operator, its origin is essentially  the same as this.

Further we remark that due to the properties of $d v$'s,
the component of matrices $W\pmb{\omega}''$, $W\pmb{\omega}'$ and $\pmb{\tau}$
in (3-14), (3-15) and (3-16) are also composed only of the data of
one or two zeroes of $P(x)$ and do not depend upon other
global structures.  In other words, for different curves expressed
by the form (3-1), the components have the same form if corresponding
zeroes are the same.

\proclaim{Lemma 3-7}\it

\roster

By introducing new coordinate in the Jacobian $\Cal J_g$,
$$
	\bold t := M^{-1} \bold u, \tag 3-19
$$
where $\bold u =(u_1, u_2,\cdots, u_g)^t$
and $\bold t :=(t_{g}, t_{g-1},\cdots, t_1)^t$,
{\it i.e.}, $u_g=t_1$,
$u_{g-1}=t_2+\mu_{g-1} t_1$, $\cdots$,
we have the relations:

\item
$$
\pi\sqrt{-1}{\pmb\omega'}^{-1}\bold u =
K(1)\bold t. \tag 3-20
$$

\item
$$
	\pmatrix \partial/\partial{u_1}\\
                 \partial/\partial{u_2}\\
                 \vdots\\
                 \partial/\partial{u_g}
         \endpmatrix
   =\ ^t M
	\pmatrix \partial/\partial{t_g}\\
                 \partial/\partial{t_{g-1}}\\
                 \vdots\\
                 \partial/\partial{t_1}.
         \endpmatrix. \tag 3-21
$$

\item
$$
\dfrac{\partial^2}{\partial
   u_i\partial u_j}
   \log \theta(u)=\sum_{kl} M_{ki} M_{lj}
     \dfrac{\partial^2}{\partial
   t_k\partial t_l}
   \log \theta(t), \tag 3-22
$$
especially,
$$
\wp_{gg}=-\dfrac{\partial^2}{\partial
   u_g\partial u_g}
   \log \theta(u)-1=-\dfrac{\partial^2}{\partial
   t_1\partial t_1}
   \log \theta(t)-1. \tag 3-23
$$

\endroster
\endproclaim

\demo{Proof}
(1) is obvious from (3-4) and (3-14). We should pay attentions on the
fixed parameters in the partial differential in (2).
$\dfrac{\partial}{\partial u_i}$ means
$\left( \dfrac{\partial}{\partial u_i}\right)_{
     u_1,\cdots,u_{i-1},u_{i+1},\cdots, u_g} $ where indices are fixed
parameters. Since we have
$$
	d t_i = \sum_{j=1}^g
 \left(\frac{\partial t_i }{\partial u_j}\right)_{
  u_1,\cdots,u_{j-1},u_{j+1},\cdots, u_g} d u_j.
\tag 3-24
$$
and
$$
     \left( \frac{\partial}{\partial u_i}\right)_{
     u_1,\cdots,u_{i-1},u_{i+1},\cdots, u_g}
          =\sum_{j=1}^g
\left(\frac{\partial t_j }{\partial u_i}\right)_{
  u_1,\cdots,u_{i-1},u_{i+1},\cdots, u_g}
\frac{\partial }{\partial t_i}. \tag 3-25
$$
Comparing (3-24) with the definition (3-19),
(2) is proved. Using (2), (3) is shown.\qed
\enddemo

Here we note that from (3-15)
$\theta(u)$ itself
vanishes but
$\tilde\theta(u):=\ee^{(\pi\sqrt{-1}\tr\pmb\tau_{\diag})/2}
\theta(u)
$
is reduced to finite summation,
$$\tilde\theta(u)
   =\sum_{n \in [-1,0]^g} \exp \left[2\pi \sqrt{-1}\left\{
    \dfrac 12 \ ^t\negthinspace (n+ \delta'')
\pmb \tau_{\off}(n+ \delta'')
    + \ ^t\negthinspace (n+ \delta'')
(\omega^{\prime -1}u+ \delta')\right\}\right].
     \tag 3-26
$$
Introducing the vector
$$
	\pmb \varepsilon
 = ((-1)^{\epsilon_1},(-1)^{\epsilon_2},
       \cdots, (-1)^{\epsilon_g} ), \tag 3-27
$$
where $\epsilon\in  \Bbb Z_2 =[0,1]$.
(3-26) can be expressed by
$$
\tilde
\theta(u)
   =\sum_{\epsilon \in \Bbb Z_2^g } \exp
\left[2\pi \sqrt{-1}\left\{
    \dfrac 18 \ ^t\negthinspace
(\varepsilon)\pmb \tau_{\off}(\varepsilon)
    + \ ^t\negthinspace (n+ \delta'')
(\omega^{\prime -1}u+ \delta')\right\}\right].
     \tag 3-28
$$

Let us summarize our results.

\proclaim{Theorem 3-8}\it

The $\tilde \wp$ for the degenerate curve, $y^2=P(x)^2 x$,
is given by
$$
  \tilde\wp_{\mu\nu}(u)=-\dfrac{\partial^2}{\partial
   u_\mu\partial u_\nu}
   \log \tilde\theta(u), \tag 3-29
$$
where
$$\tilde
\theta(u)
   =\sum_{\epsilon \in \Bbb Z_2^g }
\left[\prod_{i<j} \left(\frac{k_i- k_j}{k_i+k_j}\right)
 ^{\varepsilon_i \varepsilon_j/2}\right]
 \exp( \sum_i \varepsilon_i \xi_i(\bold t) ),\tag 3-30
$$
and $\xi_i$ is  a component of
a vector $\pmb \xi :={}^t(\xi_g,\xi_{g-1},\cdots, \xi_1)$
given by,
$$
	\pmb \xi=K(1) \bold t + \pi \sqrt{-1} \delta'.
 \tag 3-31
$$
\endproclaim

\demo{Proof}
We note that
$(\varepsilon_i\varepsilon_j+\varepsilon_j\varepsilon_i)/4
=\varepsilon_j\varepsilon_i/2$
Since we have relation,
$$
\dfrac{\partial^2}{\partial
   u_i\partial u_j}
   \log \theta(u)=
     \dfrac{\partial^2}{\partial
   u_j\partial u_l}
   \log \tilde\theta(t),\tag 3-32
$$
above theorem is obvious.\qed
\enddemo

Here we note that the coefficient of
the linear term of $\bold t$ in the exponential function in (3-30)
is replaced by a hyperelliptic integral of the second kind
in [BBEIM, I, IM1], whose integrand behaves ${\sqrt{x}}^j $ of $t_j$
around the infinite point.
 On the other hand, ours is roughly the inverse of
the integral of the first kind due to (3-20) and (3-31).
Both are connected by the Legendre relation (2-23),
$$
	{}^t\eta''\equiv 2\pi\sqrt{-1}\omega^{\prime -1} + {}^t\eta'
                 \omega^{\prime \prime}\omega^{\prime -1}.
         \tag 3-33
$$
We also note that the unnormalized differentials of the second kind (2-5)
behave around the infinite point,
$$
	d \tilde u_i|_\infty \sim \sqrt{x}^i + \text{ lower order}.
         \tag 3-34
$$
Thus our arguments and that of [BBEIM, I, IM1] are compatible.

\proclaim{Examples 3-8}
{\rm

For example, we will consider the case of genus $g=2$.
$$
\split
\tilde
\theta(u)&=-\Bigr[
 \left(\frac{k_1- k_2}{k_1+k_2}\right)^{1/2}\ee^{-\xi_1-\xi_2}
 + \left(\frac{k_1+ k_2}{k_1-k_2}\right)^{1/2}\ee^{\xi_1-\xi_2}\\
&+ \left(\frac{k_1+ k_2}{k_1-k_2}\right)^{1/2}\ee^{-\xi_1+\xi_2}
 \left(\frac{k_1- k_2}{k_1+k_2}\right)^{1/2}\ee^{\xi_1+\xi_2}
\Bigr].
\endsplit \tag 3-35
$$
Let us define
$$
\theta_\tau(t):=-
\left(\frac{k_1- k_2}{k_1+k_2}\right)^{1/2}\ee^{\xi_1+\xi_2}
\tilde\theta(u), \tag 3-36
$$
and
$$
	\xi_i':=\xi_i +\frac{1}{2} \log
\left(\frac{k_1- k_2}{k_1+k_2}\right). \tag 3-37
$$
Here the transformation (3-37) corresponds to a change of the origin
of the Jacobi variety $\Cal J_2$.
Then the new theta function $\theta_\tau$ is expressed by
$$
\theta_\tau(t)=1+\ee^{2\xi'_1}+\ee^{2\xi'_2}
           +\left(\frac{k_1- k_2}{k_1+k_2}\right)^2
                \ee^{2\xi'_1+2\xi'_2}. \tag 3-38
$$
This agrees with the tau function of two soliton solutions [H].
}
\endproclaim

\proclaim{Remark 3-10} {\rm

\roster

\item
The example 3-9 can be easily extended to arbitrary
genus $g$ case.

\item
It is easy to construct the $\tilde \theta$-function over
a curve $C_{g+1}$: $Y^2 =P(X)^2 (X-a_{g+1})^2 X$ if one uses the
coordinate $\bold t$ and knows the data
of $\tilde \theta$ over $C_g$: $y^2=P(x)^2x$ because all matrices
for $\bold t$ appearing in $\tilde\theta_t(t):=\tilde\theta(M t)$ in (3-26)
for $C_g$ are sub-matrices of corresponding ones of $C_{g+1}$.
The origin of this fact comes from the existence of $\dd v$ in (3-7) and
the degenerate type of curves $y^2=P(x)^2x$.
As long as we restrict ourselves to consider the such
simple degenerate curves, we have natural inclusion of matrices
for curves of any genera $g$ and $g'$. Then it might be easy
to construct a filter space and its inverse limit
of the filter space for set of the degenerate curves.
Then one can regard this operation for
$C_g$ and $C_{g+1}$ as a transformation in the filter space.

\item
If one wishes to extend the transformation in (2) for other type
curves, {\it e.g.} non-degenerate curves, $\tilde C_g:$ $y^2 =f(x)$ and
$\tilde C_{g+1}:$ $Y^2 =f(X) (X-a_{g+1})(X-c_{g+1})$, he will encounter some
difficulties because in general we cannot find differential forms
$\{dv_i\}$ which contains only a data of zero of $f(x)$ or
$f(X) (X-a_{g+1})(X-c_{g+1})$, even for the case of $a_{g+1}\equiv c_{g+1}$.
In other words,  these transformations mentioned in (2) does not have any
practical effect on the computations for general curves.
In fact as the similar variables to $\{dv_i\}$ of were designed in [B2 p.338],
$$
	d V_i = \gamma_i \frac{P(x)}{(x-a_i)} \frac{dx}{y}. \tag 3-39
$$
for the non-degenerated curve given in (2-1), where $\gamma_i$
is a constant.
($d V_i$ becomes $\{dv_i\}$ if the curve is degenerate like (3-1).)
It behaves as $d t_i$ for the point $(a_i,0)$ but
its global behavior are complicate like $\{du_j\}$'s and thus
Abelian integrals in terms of $d V$'s
do not have simple forms at all. It is obvious that there is no
sub-matrices structure between different curves even though their
zeroes are common like $\tilde C_g$ and $\tilde C_{g+1}$.
Comparing both differentials $d V_i$ and $\{dv_i\}$,
we can recognize how special the degenerate curve (3-1)  is.

\item Here we will comment on the derivation of Its and
Matveev [I, IM1].
They used normalized differentials of "the first kind",
$d v_i/k_i$'s; (it is noted that due to Remark 3-4,
$d v_i/k_i$'s are the third kind because in the degenerate curve (3-1),
"the first kind" here.)
In [I, IM1], noting the fact that
their differentials  are a linear combination of
the standard differentials of the "first kind"
$d u$'s with coefficients of complex numbers $\CC$,
they showed the explicit connections between $d v_i/k_i$'s
and  $d u$'s.

However they connected between $d v_i/k_i$'s and $d u$'s
by residual of computations,  using the fact that the
normalized differentials of the first kind becomes
the third kind [I, IM1].
The normalized differentials of the first kind
can be constructed by the periodic matrix $\pmb{\omega}'$ and
the standard differentials of the first kind $d u$'s
for any hyperelliptic curves. Since $d u$'s are holomorphic
 in general curves from the definition,
the residual computation has no meaning.
Thus their way is a little bit heuristic.

Further they did not comment how special the form
of $d v_i/k_i$ is.
We emphasize that it is not general that the normalized
differential of "the first kind" is expressed by only local data.
We also  emphasize the importance of the comparison
between $d v_i$'s and $dV$'s in (3-39), which
let us to know how the degenerate curve is special.

Moreover in their definitions, the coefficients of $t_1$,
$t_2$ and $t_3$ are given by the complete hyperelliptic
integrals of second kind around $\beta$'s
whereas ours are of $\omega^{\prime -1}$.
Both are connected by Legendre relation (2-23) and (3-33)
as I showed around (3-33).

\endroster

}

\endproclaim

\vskip 0.5 cm

{\centerline{\bf{ Discussion}}}

\vskip 0.5 cm

We showed that the soliton solutions are given as algebraic functions
over degenerate curves $\{y^2 = P(x)^2 x\}$, which are revised version
of work of Its and Matveev [BBEIM, I, IM1].
I believe that in physics, quantitative investigations are the most important.
 For a given problem in physics, to find an explicit solution which
can be plotted as graphs must be required, except recent elementary particle
physics.  I think that discoveries of the soliton solutions and
elliptic solutions  of nonlinear differential equations are great successes
in mathematical physics.
By virtue of them, a number of phenomena in physics become clear [L].

However from viewpoint of study of algebraic geometry, these curves expressed
by $y^2=P(x)^2 x$ are very special even in the set of hyperelliptic curves.
There are so many hyperelliptic degenerate curves which are not expressed
as $y^2 = P(x)^2 x$, {\it e.g.}, $y^2 =g(x)^2 h(x)$,
$(h(x)\not\equiv x)$ and non-degenerate curves \cite{HM}.
Even though one can regard  a curve $y^2 = P(x)^2 x$, in which the degree of
$P(x)$ is $g$, as a curve with genus $g$ from topological consideration,
it is obvious that such curve cannot represent
general properties of hyperelliptic curves.
In other words, as a function over an algebraic curve has data
of a curve itself in general, the soliton solution  exhibits
only the data of such a degenerate curve
 except very week topological property.

Further even though there is a natural inclusions as sub-matrices between
the special degenerate curves $C_g$ and $C_{g+1}$ in notations
in remark 3-10 as mentioned there, it is not clear that  the method is
some advantage for general hyperelliptic curves except formal meaning.
In other words, theories based upon special degenerate curves might not
be effective on the study of algebraic curves except formal
and weekly topological aspects \cite{HM p.42}, though
of course, it is no doubt that such considerations gave us
very fruitful and beautiful results as mentioned in the introductions.

This study shows that the soliton solutions are very special solutions.
To know the solutions space of the KdV equation requires
 concrete solutions of more general curves without undetermined parameters.
Though some mathematicians avoid quantitative considerations and their theory
might satisfy themselves, I believe that quantitative investigations
are very important at least, from physical view point, like an excellent
work \cite{HI}.
In order to do,  theories on hyperelliptic functions established in
nineteenth century [B1-3, K] are still important in this century
[BEL1-5, CEEK, EE, EEL, EEP, EGM, Ma1-2, N].

Finally, we comment on the soliton solutions of Kadomtsev-Petviashvili equation.
As it is  expected that the soliton solutions should be also constructed
from simple integrations over a degenerate algebraic curve,
it should be connected with concrete algebraic curves as we did for KdV
soliton solutions. In fact, the sigma function was also extended to
that for more general curves; the authors in [BEL5] constructed
one for the trigonal curves, which is related to Boussomesq equation.

\vskip 0.5 cm

{\centerline{\bf{ Acknowledgment}}}

\vskip 0.5 cm

I'm deeply indebted to  Prof.~Y.~\^Onishi
for leading me this beautiful theory
of Baker.
I thank Prof. V. Z. Enolskii for sending me
his interesting works and helpful comments,
and Prof. A. Its for helpful and kind comments and
telling me his works.

\centerline{\twobf Figure Caption }\tvskip

1. The homology basis in a Reimannian surface
of genus five:
$(a_i,c_i)$ and $(c,\infty)$ are the cuts
for these branching points. The $\beta$'s go
to another twin Riemannian sphere through the
corresponding cuts.

1. The homology basis in a degenerate
Reimannian surface of genus five:
$a_i$'s are singular points. The $\beta$'s go
to another twin Riemannian sphere through the
points by reparamterizations.


\Refs
\widestnumber\key{BBEIM}

%\eightptmc
\ref
  \key   {\bf {B1}}
  \by    Baker, H. F.
  \book  Abelian functions
         -- Abel's theorem and the allied theory
            including the theory of the theta functions --
  \publ  Cambridge Univ. Press
  \yr    1897, republication 1995
\endref
\ref
  \key   {\bf {B2}}
  \by    \bysame
  \paper On the hyperelliptic sigma functions
  \jour  Amer. J. of Math.
  \vol   XX
  \yr    1898
  \pages 301-384
\endref
\ref
  \key   {\bf {B3}}
  \by    \bysame
  \paper On a system of differential equations
leading to periodic functions
  \jour  Acta math.
  \vol   27
  \yr    1903
  \pages 135-156
\endref

\ref \key {\bf {BBEIM}} \by E.~D.~Belokolos, A.~I.~Bobenko, V.~Z.~Enol'skii,
  A.~R.~Its and V.~B.~Matveev \book
  Algebro-Geometric Approach to Nonlinear
  Integrable Equations \publ Springer \yr 1994 \publaddr New York \endref

\ref
  \key   {\bf {BEL1}}
  \by    Buchstaber, V. M.,  Enolskii, V. Z. and Leykin D. V.
  \paper Hyperelliptic Kleinian Functions and Application
\jour Amer. Math. Soc. Trnasl. \vol 179 \yr 1997 \pages 1-33
\endref

\ref
  \key   {\bf {BEL2}}
  \by    Buchstaber, V. M.,  Enolskii, V. Z. and Leykin D. V.
  \paper Kleinian Functions, Hyperelliptic Jacobians and Applications
  \book Reviews in Mathematics and Mathematical Physics (London)
  \eds  Novikov, S. P. and Krichever, I. M.
  \publ Gordon and Breach \publaddr India \yr 1997 \pages 1-125
\endref

\ref
  \key   {\bf {BEL3}}
  \by    Buchstaber, V. M.,  Enolskii, V. Z. and Leykin D. V.
  \paper A Recursive Family of Differential Polynomials
  Generated by the Sylvester Identity and Addition Theorems
  for Hyperelliptic Kleinian Functions
\jour Func. Anal. and Its Appl.  \vol 31 \yr 1999 \pages 240-251
\endref

\ref
  \key   {\bf {BEL4}}
  \by    Buchstaber, V. M.,  Enolskii, V. Z. and Leykin D. V.
  \paper Rational analogues of the ablelian functions
\jour Func. Anal. Appl.  \vol 33 \yr 1999 \pages 1-15
\endref


\ref
  \key   {\bf {BEL5}}
  \by    Buchstaber, V. M.,  Enolskii, V. Z. and Leykin D. V.
  \paper Uniformization of the Jacobi varieties of
trigonal curves and nonlinear differential equations
\jour Func. Anal. Appl.  \vol 34 \yr 2000 \pages 1-15
\endref



\ref
  \key   {\bf {CEEK}}
  \by    Christiansen, P. L., Eilbeck, J. C.,
 Enolskii, V. Z. and Kostov, N. A.
  \paper Quasi-periodic and periodic solutions
for coupled nonlinear Schr\"odinger equations of
Manakov type
\jour Proc. R. Solc. Lond. A \vol 456 \yr 2000 \pages
2263-2281
\endref


\ref
  \key   {\bf {DKJM}}
  \by    Date, E., Kashiwara, M., Jimbo, M.  and Miwa, T.
  \paper Transformation groups for soliton equations,
  \book Nonlinear Integrable Systems - Classical Theory and
        Quantum Theory
  \eds  Jimbo, M. and Miwa, T.
  \publ World Scientific \publaddr Singapore \yr 1983
  \pages 39-119
\endref


\ref \key {\bf {EGM}}
\by Edelstein J. D., G\'omez-Reino M. and
Mari\~no, M.
\paper
Blowup formulae in Donaldson-Witten theory and
integrable hierarchies
\jour
Advances in Theoretical and Mathematical Physics
\vol 4 \yr 2000 \pages 503-543
 \lang
hep-th/0006113
\endref

\ref
  \key   {\bf {EE}}
  \by    Eilbeck, J. C. and Enolskii, V. Z.
  \paper Bilinear operators and the power series for the
       Weierstrass $\sigma$ function
  \jour J. Phys. A: Math. \& Gen.
 \pages 791-794 \yr 2000 \vol 33
\endref


\ref
  \key   {\bf {EEL}}
  \by    Eilbeck, J. C.,  Enolskii, V. Z. and Leykin D. V.
  \paper On the Kleinian Construction of Abelian
         Functions of Canonical Algebraic Curves
  \book Proceedings of the Conference SIDE III:
        Symmetries of Integrable Differences Equations,
        Saubadia, May 1998, CRM Proceedings and Lecture Notes
 \pages 121-138 \yr 2000
\endref

\ref
  \key   {\bf {EEP}}
  \by    Eilbeck, J. C.,  Enolskii, V. Z. and Perviato E.
  \paper Varieties of elliptic solitons
  \jour J. Phys. A: Math. Gen
  \yr 2000 \vol 456 \pages 2263-2281
\endref

\ref
  \key   {\bf {GGKM}}
  \by    Gradner, C. S.m Greene, J. M., Kruska, M. D.,
         and Miura, R. M.
  \paper Method for solving the Korteweg-de Vries equation
  \vol   19
  \yr    1967
  \pages 1095-1097
  \jour  Phys. Rev. Lett.
\endref


\ref
  \key   {\bf {H1}}
  \by    Hirota, R.
  \paper Direct Method of Finding Exact Solutions of
         Nonlinear Evolution Equations
  \book B\"acklund Transformations,
         Lecture Notes in Math. 515
  \eds R. M. Miura
  \publ Springer
  \publaddr Berlin
  \yr    1976
\endref
\ref
  \key   {\bf {H2}}
  \bysame
  \paper Exact Solutions of Korteweg-de Vries Equation for
         Multiple Collisions of Solitons
  \jour Phys. Rev. Lett.
  \yr    1971 \vol 27 \pages 1192--1194
\endref
\ref
  \key   {\bf {HI}}
  \by    Hirota, R. and Ito, M.
  \paper A Direct Approach to Multi-Periodic Wave Solutions
         to Nonlinear Evolution Equations
  \jour J. Phys. Soc. Jpn
  \vol   50
  \yr    1981
  \pages 338-342
\endref
\ref
  \key   {\bf {HM}}
  \by    Harris, J. and Morrison, I.
  \book  Moduli of Curves
  \publ  Springer GTM 187
  \publaddr Berlin
  \yr    1998
\endref
\ref
  \key   {\bf {I}}
  \by     Its A.R.
  \paper On Connections between Solitons and
         Finite-Gap Solutions of the
         Nonlinear Schr\"odinger Equation
  \jour Sel. Math. Sov.,
  \vol   5
  \yr    1986
  \pages 29-43
\endref
\ref
  \key   {\bf {IM1}}
  \by    Its A.R., and   Matveev V.B.
  \paper On a Class of Solutions
    of the Korteweg-de Vries
    Equation
  \jour Problemy Mat. Fiz.,
  \vol   8
  \yr    1976
  \pages 70-92
  \lang Russian
\endref

\ref
  \key   {\bf {IM2}}
  \by    Its A.R., and   Matveev V.B.
  \paper  Schr\"odinger operators
     with finite-gap spectrum and N-soliton solutions of
     the Kortweg-de Vries equation
   \jour Theoret. Math. Phys.
  \vol   23
  \yr    1975
  \pages 343-355
\endref

\ref
  \key {\bf {Kl}}
  \by Klein F
  \yr 1886
  \book Ueber hyperelliptische Sigmafunctionen
\jour Math. Ann. \vol 27 \pages 431-464
\endref

\ref \key {\bf {Kr1}}
     \by Krichever I. M.
      \paper Methods of algebraic geometry in the theory
      of nonlinear equations
       \jour Russian Math. Surverys
       \vol 32 \pages 185-213 \yr 1977 \endref

\ref \key {\bf {Kr2}}
     \bysame
      \paper Foreword in republication of [B1] \endref
\ref
  \key {\bf {L}}
  \by Lamb, G. L., Jr.
  \yr 1980
  \book Elements of Soliton Theory
  \publaddr New York
  \publ Wiley-Interscience
\endref

\ref\key   {\bf {Ma1}}
  \by Matsutani, S.
      \paper Hyperelliptic Solutions of KdV and
KP equations:Reevaluation of Baker's Study on
Hyperelliptic Sigma Functions
      \jour nlin.SI/0007001
        \endref
\ref
  \key {\bf {Ma2}}
  \bysame
      \paper Closed Loop Solitons and Sigma Functions: Classical and
      Quantized Elasticas with Genera One and Two
        \jour J. Geom. Phys. \yr 2001 \vol 698
\endref

\ref\key   {\bf {Ma3}}
         \paper Explicit Hyperelliptic Solutions of
Modified Korteweg-de Vries Equation:
Essentials of Miura Transformation
          \jour preprint, nlin.SI/0108002 \yr 2001
        \endref


\ref
  \key   {\bf {N}}
  \by Nijhoff, F. W.
      \paper Discrete Dubrovin Equations and Separation of
Variables for Discrete Systems
   \jour Chaos, Solitons and Fractals \vol 11
\yr 2000 \pages 19-28     \endref

%\ref
%   \key{\bf {NMPZ}}
%   \by Novikov, S., Manakov, S. V., Pitaevskii, L. P.,
%       and Zakharov, V. E.
%    \yr 1984
%   \book Theory of Solitons: The Inverse Scattering Method.
%   \publaddr New York
%   \publ Consultants Bureau
%\endref



\ref \key {\bf{\^O1}} \by \^Onishi Y. \paper Complex
multiplication formulae for curves of genus three
\jour Tokyo J. Math. \vol 21 \pages 381-431 \yr1998
\endref
\ref \key {\bf{\^O2}}\bysame \paper Determinatal Expressions
for Some Abelian Functions in Genus Two
\jour preprint \yr2000
\endref

\ref
  \key   {\bf {SN}}
  \by    Sato, M, Noumi
      \paper Soliton Equations and Infinite
        Dimensional Grassmann Manifold
      \jour Sofia University Lecture Note
      \vol 18 \yr 1984
       \lang Japanese
       \publaddr Tokyo
     \endref
\ref
  \key   {\bf {SS}}
  \by    Sato, M and Sato, Y
      \paper Soliton Equations as Dynamics Systems on Infinite
        Dimensional Grassmann Manifold
      \book Nonlinear Partial Differential Equations in Applied
       Science
      \eds Fujita, H, Lax, P.D. and Strang, G
       \publ Kinokuniya/North-Holland
       \publaddr Tokyo
      \yr 1984
     \endref



\endRefs
\enddocument